%% file: type1.tex
\documentclass[12pt]{article}
\usepackage{epsfig}
\catcode`@=11
\def\marginnote#1{}
\newcount\hour
\newcount\minute
\newtoks\amorpm
\hour=\time\divide\hour by60
\minute=\time{\multiply\hour by60 \global\advance\minute by-\hour}
\edef\standardtime{{\ifnum\hour<12 \global\amorpm={am}%
        \else\global\amorpm={pm}\advance\hour by-12 \fi
        \ifnum\hour=0 \hour=12 \fi
        \number\hour:\ifnum\minute<10 0\fi\number\minute\the\amorpm}}
\edef\militarytime{\number\hour:\ifnum\minute<10 0\fi\number\minute}
\def\draftlabel#1{{\@bsphack\if@filesw {\let\thepage\relax
   \xdef\@gtempa{\write\@auxout{\string
      \newlabel{#1}{{\@currentlabel}{\thepage}}}}}\@gtempa
   \if@nobreak \ifvmode\nobreak\fi\fi\fi\@esphack}
        \gdef\@eqnlabel{#1}}
\def\@eqnlabel{}
\def\@vacuum{}
\def\draftmarginnote#1{\marginpar{\raggedright\scriptsize\tt#1}}
\def\draft{\oddsidemargin 0.0truein
        \def\@oddfoot{\sl preliminary draft \hfil
        \rm\thepage\hfil\sl\today\quad\militarytime}
        \let\@evenfoot\@oddfoot \overfullrule 3pt
        \let\label=\draftlabel
        \let\marginnote=\draftmarginnote
   \def\@eqnnum{(\theequation)\rlap{\kern\marginparsep\tt\@eqnlabel}%
\global\let\@eqnlabel\@vacuum}  }
\def\slash#1{\setbox0=\hbox{$#1$}#1\hskip-\wd0\dimen0=5pt\advance
       \dimen0 by-\ht0\advance\dimen0 by\dp0\lower0.5\dimen0\hbox
         to\wd0{\hss\sl/\/\hss}}
\catcode`@=12
\newcommand{\nn}{\nonumber}

\newcommand{\be}{\begin{equation}}
\newcommand{\ee}{\end{equation}}
\newcommand{\bea}{\begin{eqnarray}}
\newcommand{\eea}{\end{eqnarray}}

\newcommand{\gev}{ {\rm GeV} } 
\newcommand{\tev}{ {\rm TeV} } 
\newcommand{\mm}{ {\rm mm} } 
\newcommand{\cm}{ {\rm cm} } 
\newcommand{\mim}{ {\mu \rm m} }

\begin{document}
\pagestyle{empty}
%
%
\vskip 1cm
\centerline{\Large \bf Anisotropic Type I String Compactification, }
\vskip 0.3cm
\centerline{\Large \bf Winding Modes and Large Extra Dimensions}
\vskip 1.5cm
\centerline{\bf \Large A. Donini\footnote{andrea.donini@roma1.infn.it}
                   and S. Rigolin\footnote{stefano.rigolin@pd.infn.it} } 
\vskip 0.3cm
\begin{center}
\small{
Departamento de F\'{\i}sica Te\'orica C-XI, \\
Universidad Aut\'onoma de Madrid, 
Cantoblanco, 28049 Madrid, Spain.} 
\end{center}
\vskip 1truecm
\begin{center}
\small{\bf \large Abstract} 
\\[7mm]
\end{center}
%
%
\begin{center}
\begin{minipage}[h]{14.0cm}
We discuss the structure of general Anisotropic Compactification in 
Type I $D=4$, $N=1$ string theory. It is emphasized that, in this context, 
a possible interpretation of $M_{Planck}$ as ``dual'' to (at least) one of the 
Kaluza-Klein or Windings Modes could provide interesting interpretations 
of the ``physical scales'' and the GUT coupling. Some of the scenarios 
presented here are strictly connected with the phenomenological proposal 
of TeV-scale gravity (i.e. millimiter compactification). We show that 
in this scenario new and probably dominant effects should arise from the 
presence of ``low-energy'' Winding Modes of usual SM particles. 
Stringent bounds on the Planck mass in $4+n$ dimensions are derived
from the existing experimental limits on massive replicas of SM gauge bosons. 
Non-observation of Winding Modes at the planned accelerators could provide
stronger bounds on the $4+n$ dimensional Planck mass than those from 
graviton emission into the bulk. Some comments on 
other possible phenomenological interesting scenarios are addressed.
\end{minipage}
\end{center}
PACS: 11.10.Kk; 04.50.+h; 11.25.Mj \\
Keywords: Type I String theory; String Phenomenology; Extra Dimensions. 
%
%
\vskip 0.8cm
\vfill\eject
\newpage
\pagestyle{plain} 
\setcounter{page}{1}
\setcounter{footnote}{0}
%
%
\section*{Introduction}
\label{introduction}
%
%
The key to the recent development of string theory was the discovery of duality 
symmetries. Dualities not only relate the strong and weak coupling 
limits of different string theories, but also suggest a way to compute 
certain strong coupling results in one string theory by mapping it to 
weak coupling result in a dual one. 

For many years mostly all the attention was devoted to the theoretical 
and phenomenological analysis of the $E_8 \times E_8$ weak coupled heterotic 
string theory. But this theory, beyond many successful (and not trivial) 
predictions, gives rise to a gauge coupling unification scale that is 
in contrast with the usual GUT value extrapolated from low-energy data. 
In fact the $E_8 \times E_8$ weak coupled heterotic string theory make a 
definite prediction of the relation between the gravitational scale 
($M_{Planck}$), the string scale $M_s$ and the GUT coupling. From this 
relation it comes out that either the scale where gravity becomes of the 
same order of gauge interactions (the string unification scale) should be 
higher respect to the gauge unification scale, $M_{GUT}$, either the 
$M_{Planck}$ should be smaller than the usual value ($ \approx 10^{19}$). 
Various proposal have been thought for dealing with this problem in the 
context of perturbative heterotic string theory, but none of them is really 
compelling (see \cite{dienes} for a comprehensive review on the subject). 

An alternative approach was proposed by Witten in \cite{witten}, where 
it is proposed to look for a solution to the GUT-gravity unification 
problem in the strongly coupled regime. When string coupling is strong 
large corrections can appear that substantially modify the relation between 
the gravitational, string and gauge couplings. 
Using the new duality symmetries, it has been shown that the strongly 
coupled regime of the $SO(32)$ heterotic is described by the weakly 
coupled Type I string theory \cite{polwit}, while the for the $D=10$ 
$E_8 \times E_8$ heterotic the strong coupling limit is given by the 
weakly coupled $D=11$ M-theory compactified on $S^1/Z_2$ \cite{horwit}.
The new features of these schemes is that now the string scale is no more 
fixed to a particular value, and so, why not, just behind the corner. 
On the other hand, the Planck scale, no more connected directly to the 
string scale by a definite relation, seems to loose its fundamental role 
as a physical scale and appears to be either imposed by hand either 
accidental. In a recent paper, \cite{biq}, $M_{Planck}$ was suggested 
to be ``dual'' to the EW scale, having in this way a motivation. 
However, in the context of Isotropic Compactification in Type I string 
theory both scales (although connected) are still external input. 
We suggest that a ``physical motivation'' of the $M_{Planck} - M_{EW}$ 
``duality'' could be recovered in the context of Anisotropic Compactification.

Recently, an interesting scenario of low-energy quantum gravity has been 
proposed \cite{lykken,add}. Its phenomenological interest relies on the fact 
that future accelerator and gravity experiments could in principle observe 
some effects due to Large Extra Dimensions with a compactification 
radius at mm. This scenario can be recovered in the 
framework of Type I string theory, as proposed in \cite{aadd}.
However, particularities of the specific type of compactification 
implemented could in general strongly modify the phenomenological 
impact of this hypothesis, changing by many orders of magnitude
the strenght of the typical signatures. In particular, 
adopting a naive Isotropic Compactification model, 
graviton emission in the extra dimensions is strongly suppressed 
with respect to the favoured case of Anisotropic 
Compactification with only two large extra dimensions \cite{grw}.
In this paper we show that, however, in a comprehensive treatment
of Anisotropic Compactification in Type I strings, quite generally, 
winding modes can appear below the string scale for the typical values 
currently quoted in the literature. We claim that the winding modes 
could represent then a possible dominant string effect accessible 
to present or near future experiments. 

In sect.~\ref{type1} we shortly remind some issues related to Type I string 
theory and D$p$-branes. We also give the general relations for the 
gravitational, string and gauge couplings, obtained after compactification 
on a six-dimensional Calabi-Yau manifold both for Isotropic and Anistropic 
Compactification. 

In sect.~\ref{nostro} we consider the simple assumption
that the Planck scale could be ``dual'' to the lowest scale in the model
(being it a compactification scale or a winding mode) and describe
qualitatively three typical scenarios, where $M_s = 10^3, 10^{11}$ and 
$10^{16}$ GeV.

In sect.~\ref{winLED} we analyse the low-energy quantum gravity scenarios 
studied in the recent literature. We stress the new phenomenological 
aspects due to presence of low-energy winding modes and their relation 
with the large compactification radius and with the D-dimensional 
Planck mass.

Eventually, in sect.~\ref{concl} we draw our conclusions.

%
\input{tychap1.tex}

%
\input{tychap2.tex}

%
\input{tychap3.tex}

%
\section{Conclusions}
\label{concl}

The Anisotropic Compactification scenario of Type I string theory
represent the preferred framework for TeV-scale gravity models. 
However, in this scenario a large number of mass scales naturally appear 
possibly reintroducing the hierarchy problem and suggesting that a 
stabilization mechanism of the compactification radii is at work. 
Moreover, in Type I strings, the string scale is not fixed
to a particular value and the Planck mass seems to loose 
its fundamental role. This situation has to be compared to that
of heterotic string theory. In this paper, we considered 
$M_{Planck}$ to be ``dual'' to the lowest mass scale of the model. 
Within this simple assumption, all the mass scales can be
deduced by only two free parameters, the gauge coupling $\alpha_3$
and one mass scale (for example $M_s$). Moreover, the Planck scale 
recovers a physical meaning. In sect.~\ref{nostro} we analysed 
qualitatively in the context of our Ansatz the different scenarios 
that arise for three reference values of $M_s = 10^3, 10^{11}$ 
and $10^{16} \ \gev$. We noticed that quite generally winding modes 
could appear below the string scale, thus being of some phenomenological
relevance. 

In the case of $n=2$ Large Extra Dimension, direct searches of the lightest 
winding modes can put stringent bounds on both the largest compactification scale, 
$R_1$, and the six-dimensional Planck mass, $M_{(6)}$.
We stress that these results are completely independent from our Ansatz
and follows only by assuming Type I string theory.
With the present data on massive replicas of SM gauge bosons, 
$M_{Z^\prime_{SM}} \ge 700 \ \gev$, we obtain $M_{(6)} \ge 1.8 \ \tev$ 
and $R_1 \le 0.15 \ \mm$ for a reasonable value of $\alpha_3$.
Search for direct production of winding modes at LHC or NLC
can then represents a useful tool to explore the parameter space of 
Large Extra Dimension models derived from Type I string theory
and give a possibly cleaner signature than graviton emission into the bulk. 

The phenomenology of winding modes is quite interesting and 
deserves a careful study.

%
%
\section*{Acknowledgements}
We kindly thank M. B. Gavela, L. Ib\'a\~nez and C. Mu\~noz 
for continuous suggestions and discussions during the completion of this work.
We also thank A. De Rujula, F. Feruglio and F. Zwirner
for useful comments. 
A. Donini acknowledges the I.N.F.N. for financial support. 
S. Rigolin acknowledges the European Union for financial support 
through contract ERBFMBICT972474.
%

%
\end{document}

%% file: tychap1.tex
%
\section{Type I String}
\label{type1}
%
%
Type I string models are widely used as a framework for low-energy quantum 
gravity theories. Since our main motivation is of phenomenological nature, 
in this section we shortly remind some issues related to Type I string theory 
and D$p$-branes relevant to this paper. For a deeper and more formal
sight into the subject see refs.~\cite{polrev,rev}.
%
%
\subsection{New features of Type I strings: D$p$-branes}
D$p$-branes are classical solutions that appear in many string 
theories. They can be thought as extended objects with $p$ spatial 
dimensions and localized in all the other spatial directions. 
In the weak string coupling limit, D$p$-branes can be understood as static 
surfaces on which Type I open strings can end. In the ``old-fashioned view'', 
Type I open strings were free to move in all the 10D space-time (a 9-brane), 
i.e. they had only Neumann boundary conditions. 
It has been noticed \cite{polrev} that in the compactification procedure 
new sets of D$p$-branes with $p < 9$ can appear. As a consequence, Type I 
open strings have Dirichlet boundary conditions in the $9-p$ coordinates 
transverse to the D$p$-brane and Neumann boundary condition in the brane
spatial directions. This means that their ends can freely move only in the 
brane $(p+1)$ dimensions. To each set of D$p$-branes is associated a
gauge group. We assume that SM particles are massless excitations (zero 
modes) of open strings starting and ending on the same brane\footnote{Only 
strings starting and ending on the same D$p$-brane or in different branes 
with a non vanishing intersection of the world-volume can have massless 
zero modes.}. Consequently, gauge and matter SM fields are tied to the brane 
and can not see the transverse extra dimensions. 

In Type I string theory also a closed string sector is present. Closed 
strings are singlets respect to the gauge group and represent the 
gravitational interaction. They are free to propagate in the full 10D space 
can mediate SUSY or EW breaking occuring in a brane sector not coincident 
with the SM one.

In the simple case of toroidal compactification $T^2 \times T^2 \times T^2$,
we have three compactification radii $R_i (i = 1, 2, 3)$ associated
to the three complex dimensions $X_i$. Strings have two kind of excitations: 
Kaluza-Klein (KK) and winding modes. 
The KK modes have masses typically of the order of the 
compactification scales, $M_i = 1 / R_i$, whereas the masses of the winding 
modes are $M_{\omega_i} = M_s^2/M_i$, with $M_s$ the string scale. 
The mass squared of a given excitation is:
\be
M^2(m_i, n_i) = M_0^2 + \sum_{i=1}^3 \left( m_i^2 M^2_i + n_i^2 M_{\omega_i}^2 \right) \ ,
\qquad m_i, \  n_i = 0, \pm 1, \pm 2, \dots
\ee  
where $M_0$ stands for other $R_i$-independent contributions to the mass
and $(m_i, n_i)$ are the momentum and winding numbers. 
%
%
\begin{table}[t]
\centering
\begin{tabular}{||c|c|c|c|c|c||}
\hline\hline
         & 3-brane & $5_1$-brane & $7_1$-brane & 9-brane & gravity \\
\hline\hline
& & & & &\\
KK       & none    & $M_1$ & $M_2, M_3$  & $M_1,M_2,M_3$ &  $M_1,M_2,M_3$\\
& & & & &\\
\hline
& & & & &\\
Windings & $M_{\omega_1}, M_{\omega_2}, M_{\omega_3}$ & 
$M_{\omega_2}, M_{\omega_3}$ & 
$M_{\omega_1}$ & none &  $M_{\omega_1}, M_{\omega_2}, M_{\omega_3}$ \\
& & & & & \\
\hline \hline
\end{tabular}
\caption{\it{
String modes felt by the gauge group living on a specific D$p$-brane 
(we consider for simplicity just $5_1$- and $7_1$-branes) and by gravity.}}
\label{tabmode}
\end{table}
%
%
In Type I models, closed strings have both KK and winding excitations, 
whereas open strings starting and ending on a D$p$-brane can only have 
KK modes associated with the $p-3$ complex dimensions belonging to the brane 
and winding modes related to the $9-p$ directions perpendicular to 
the brane. For example, a $5_1$-brane has one infinite tower of KK modes
associated to the brane internal complex dimension $X_1$ and two infinite 
towers of winding modes associated to the two external ones, $X_2, X_3$;
three different $5_i$-branes are possible by taking as internal dimension $X_i$.
A $7_1$-brane has two KK modes associated to the two internal complex dimensions 
$X_2, X_3$ and one winding mode associated to the external dimension $X_1$;
also in this case, three different $7_i$-branes are possible.
In Tab.~\ref{tabmode} we schematically remind which kind of string 
excitations are felt by the gauge group confined on a particular D$p$-brane 
and by gravity. 
%
\subsection{Compactification in Type-I strings}
%
The relevant terms in the D=10 $N=1$ effective low-energy action appearing 
in Type I string theory are \cite{witten}:   
\be
S_{10} \ =\ - \int {{d^{10}x} \over {(2\pi )^7}}
 \sqrt{-g}\  \left(\  {{ M_s^8}\over {\lambda^2}} \ R \ 
+\  {{M_s^6}\over {\lambda}} \ {1\over 4} F^2_{(9)} \ +\ ...\  \right)\ ,
\label{d10}
\ee
where $\lambda$ is the string coupling, $M_s$ 
is the string scale, $R$ is the trace of the 10D Ricci tensor and $F_{(9)}$ 
is the gauge field strength associated to the 9-brane sector. 
By dimensional reduction to D=4 on an orbifold with underlying 
six-dimensional compact torus $T^2 \times T^2 \times T^2$, one obtains:
\bea
S_{4} \ &  = & \ - \int {{ d^4 x}\over {2\pi }}
\sqrt{-g}\ \left(\ {{R_1^2R_2^2R_3^2 M_s^8}\over{\lambda^2}} \ R \ + \  
{{R_1^2R_2^2R_3^2M_s^6}\over{\lambda}} \ {1\over 4} F^2_{(9)}\right. \nn \\ 
& + & \left. \ \sum _{i\not= j\not=k\not= i} {{R_i^2 R_j^2 M_s^4}\over 
{\lambda}} \ {1\over 4} F^2_{(7_k)} \ + \ \sum_{j=1}^3 
{{R_j^2 M_s^2}\over {\lambda}} \ {1\over 4}  F^2_{(5_j)} \ +
{1\over {\lambda}}\ {1\over 4}  F_{(3)}^2  \ +\ ...\  \right)\ ,
\label{d04}
\eea
where the compactified volume is defined as $V = \Pi_i (2 \pi R_i)^2$.
In eq.~(\ref{d04}) we displayed all the possible kinetic terms for gauge bosons that can 
in general appear coming from different D$p$-branes ($p=9,7,5,3$) sectors. 
If N=1 SUSY is to be preserved one should consider, in a specific model, 
only branes satisfying the constraint $|p - p'|=0,4$, with $p, p'$ the spatial 
dimensions of two given D$p$-branes. From eq.~(\ref{d04}) 
one recovers the following relation between the gravitational constant 
$G_N$, the string coupling $\lambda$, the string scale of the theory 
$M_s$ and the scales of the compactified dimensions $M_i$ 
\cite{witten,imr}:
\be
G_N \ = \ {1 \over {M_{Planck}^2}} \ =\  {{\lambda^2 M_1^2 M_2^2 M_3^2} 
       \over {8 M_s^8}} \ .
\label{planck}
\ee
The gauge couplings on the different branes present in the theory are 
related to the string scale and the compactification scales by:
\bea
\alpha_9 \ = \   {{\lambda M_1^2 M_2^2 M_3^2} \over {2 M_s^6}} \ &;& 
\qquad  \alpha_{7_i} \ = \ {{\lambda M_j^2 M_k^2} \over {2 M_s^4}} \ 
\ , i \not=j \not=k \not=i  \nn \\
\alpha_{5_i} \ = \ {{\lambda M_i^2} \over{2 M_s^2}} \ &;& \qquad 
\alpha_3 \ = \ {{\lambda}\over 2} \ . 
\label{gaugci}
\eea
From eq.~(\ref{planck}) we derive the perturbative bound on the string 
coupling,
\be
\lambda = 2 \sqrt{2} \frac{M_s^4}{M_1 M_2 M_3 M_{Planck} } \le {\cal O}(1),
\label{lambda}
\ee 
that defines the range of validity of the effective action in eq.~(\ref{d10}). 
Substituting eq.~(\ref{lambda}) in eq.~(\ref{gaugci}) the gauge couplings, 
for the different D$p$-brane sectors, can be directly related to the 
relevant masses of the theory by:
\bea
\frac{\alpha_9 M_{Planck}}{\sqrt{2}} = \frac{M_1 M_2 M_3}{M_s^2} \ &;& 
\qquad \frac{\alpha_{7_i} M_{Planck}}{\sqrt{2}} = \frac{M_j M_k}{M_i} 
\ , i \not=j \not=k \not=i  \nn \\
\frac{\alpha_{5_i} M_{Planck}}{\sqrt{2}} = \frac{M_i M_s^2}{M_j M_k} \ 
\ , i \not=j \not=k \not=i  &;& 
\qquad \frac{\alpha_3 M_{Planck}}{\sqrt{2}} = \frac{M_s^4}{M_1 M_2 M_3} \ .
\label{gaugcm}
\eea
In the case of Isotropic Compactification \cite{biq,imr}, where all 
compactification radii are taken to be equal, $R_i = 1/M_i = 1/M_c$, the 
previous formulae can be simplified to:
\be
\frac{\alpha_p M_{Planck}}{\sqrt{2}} = \frac{ M_c^{(p-6)}}{M_s^{(p-7)} }
\label{gaugcmi}
\ee
with $p$ labelling the different D$p$-branes. 

From eq.~(\ref{gaugcm}) the meaning of T-duality, intuitively defined by 
the relation
\be
T_i \ : \qquad M_i \to M_{\omega_i} = \frac{M_s^2}{M_i} \ ,
\label{tdual}
\ee
appears evident. For example it is immediate to see how the $T_1$ duality 
connects the 9-brane with the $7_1$-brane sector or the $5_1$-brane with 
the $3$-brane one. These sets of relations are somewhat redundant: since 
different D$p$-branes are related by $T$-duality, the formulae for the 
gauge couplings are not at all independent. Starting from, say, the 
3-brane gauge coupling the others can be derived by interchanging the 
compactification/winding scales in the appropriate way. As our discussion 
will be principally based on eqs.~(\ref{gaugcm},\ref{gaugcmi}) it is 
obvious that we can choose a reference set of D$p$-brane, for example the 
3-brane one, and then all the conclusions for other sets of D$p$-branes can 
be obtained via T-duality transformations. 


%% file: tychap2.tex
\section{Anisotropic Compactification Scenarios}
\label{nostro}
%
In the previous section we recalled the general formulae that relate 
the gauge coupling of a specific D$p$-brane sector and the Newton constant 
with the string and the compactification masses. We now describe in more 
details the Anisotropic Compactification scenarios and try 
to sketch out possible phenomenological implications. Connections with 
present and future gravity and accelerator experiments will be postponed 
to sect.~\ref{winLED}. 

Isotropic Compactification scenarios and possible phenomenological 
consequences have been deeply investigated in \cite{biq,imr}. While a lot 
of attention is devoted in literature to the phenomenological analyses of 
Large Extra Dimensions, a clear systematization of the Anisotropic 
Compactification in Type I string, that it is the preferred framework for 
such models, is still lacking. We try to partially fill this gap in the 
present section.

The simplest ``phenomenological'' scenario is obtained embedding the 
SM gauge group $SU(3)_C \times SU(2)_L \times U(1)_Y$ in only one brane 
sector\footnote{We are not considering, for simplicity, the case in which 
the SM gauge group is embedded in different D$p$-branes. For a discussion 
of this case see \cite{imr} where some examples are analyzed and possible 
problems arising from the measured value of $\sin^2 \theta_W$ are discussed.}.
Let's choose for definiteness the 3-brane one. As stated before, the 
discussion will follow exactly the same, using the appropriate T-dualities, 
if one would choose another D$p$-brane embedding. However, this 
scenario offers the best intuitive approach to phenomenology. In fact, we 
end up with closed strings (gravity) propagating in the full 10D space and 
open strings (gauge interactions) tied to 3-branes. The SM low-energy 
fields are thus confined on the 3-brane sector (the usual 4D space-time) 
and cannot propagate in the six extra dimensions. As appears from 
Tab.~\ref{tabmode} the SM gauge group can see only winding modes excitations.
SUSY and/or EW breaking can be transmitted to the observable sector from 
other D$p$-brane sectors, separated from the SM one, via gravitational 
interaction or ``massive'' string excitation between the branes.

In the crowded zoology of Anisotropic scenarios, we consider explicitly 
the case of two identical compactification scales\footnote{The extension 
to the case with three different compactification scales does not add any 
new interesting physical problematic and so we do not analyse it.} $M_2 = 
M_3 = M_c$ and with the third scale set $M_1 \neq M_c$. Hence, eq.~(\ref{gaugcm}) 
in terms of the KK and winding masses reads:
\be 
\frac{\alpha_3 M_{Planck}}{\sqrt{2}} 
    = \left( \frac{M_s^4}{M_1 M_c^2} \right) 
    = \left(\frac{M_{\omega_c}}{M_c}\right) M_{\omega_1} 
    = \left(\frac{M_{\omega_1}}{M_c}\right)^2 M_1 \ . 
\label{gaugnos}
\ee
Consequently, the string scale is no more fixed to any particular value 
and the Planck scale, no more connected directly to the string scale by a 
definite relation, seems to loose its fundamental role as a physical scale 
and appears to be either imposed by hand or accidental. 
In principle, any two of the three scales on the right hand side of eq.~(\ref{gaugnos}) 
can assume arbitrary large/small values (if not excluded by experimental bound). 
This situation has to be compared, for example, with the heterotic prediction:
\be
\frac{ \sqrt{\alpha_{GUT}} M_{Planck} }{2} = M_s 
\label{gaughet}
\ee
where the string scale is fixed to a definite value, $M_s \sim 10^{18}$.

In order to re-establish a role for the Planck scale, and to 
reduce the number of free parameteres of the model, 
we make the following Ansatz: 
\begin{quote} 
{\bf Ansatz:} \newline
\it
The Planck scale is ``dual'' to the smallest mass scale present in the 
theory, being it a compactification scale $M_1$ or its winding mode 
$M_{\omega_1}$, i.e.:
\be
M_1 \qquad {\rm or } \qquad M_{\omega_1} = \frac{M_s^2}{M_{Planck}} \nn \ .
\label{ansatz}
\ee
\end{quote}
Within this simple assumption, all the mass scales of the 
theory are derived by only two input parameters: the gauge coupling 
$\alpha_3$ and one mass, either the lowest mass, $M_1$ ($M_{\omega_1}$) 
or the string scale $M_s$. Moreover, a new physical meaning for the
Planck scale is recovered, being it the largest mass scale in the game.
The equivalence between the Planck mass and $M_1$ or $M_{\omega_1}$
should descend by some underlying fundamental principle,
that is beyond the motivation of this paper.

In general, two different scenarios will appear:
\begin{enumerate}
\item The smallest scale of the theory is the compactification scale $M_1$. 
So the Planck scale is the related winding mode mass, $M_{Planck} = 
M_{\omega_1}$, and for the gauge coupling we get:
\be
\label{ratio1}
\frac{\alpha_3}{\sqrt{2}} = \left( \frac{M_{\omega_c}}{M_c} \right) \ .
\ee
The mass relations for this scenario are depicted in Fig.~\ref{lowkk}.
Below $M_1$ gravity and the SM live in four dimensions. At $M_1$, 
gravity sees two extra dimensions and the graviton acquire an infinite tower
of KK modes. The SM particles are still living in 4D, since they do not
feel KK excitations, as was explained in sect.~\ref{type1}.
However, at $M_{\omega_c}$ both gravity and gauge interactions feel
the effect of extra dimensions, since both are sensitive to winding
mode excitations. Hence, gravity feels the full 10D space-time
whereas SM feels 4 extra dimensions and the SM particles
acquire an infinite tower of winding modes. At $M_s$, string unification 
occurs: the gravitational constant become of the same order of magnitude 
of the gauge couplings. Above the string scale 10D field theory must be 
replaced by string theory.

\item If the smallest scale of the theory is the winding mode $M_{\omega_1}$,
then $M_{Planck} = M_1$ and we have:
\be
\label{ratio2}
\frac{\alpha_3}{\sqrt{2}} = \left ( \frac{M_{\omega_1} }{M_c } \right )^2 \ .
\ee
This case is illustrated in Fig.~\ref{lowwind}, and the meaning of the 
different scales follows what already explained in the previous case.
\end{enumerate}

The main simplification arising from our Ansatz is that the whole 
mass scales pattern depend on two parameters only. To illustrate this
statement, by taking as input parameter $M_1$ and $\alpha_3$, 
we derive in the first scenario:
\bea
M_s          &=& \sqrt{ M_1 M_{Planck} } \nn \\
M_c          &=&  \left ( \frac{\sqrt{2} }{\alpha_3} \right )^{1/2} M_s \nn \\
M_{\omega_c} &=&  \left ( \frac{\alpha_3}{\sqrt{2}} \right )^{1/2} M_s \nn
\eea
We see that a simple {\it accordion picture} of all the scales in the 
model can be drawn.

Eventually, we notice that the gauge coupling 
has a natural explanation in terms of the geometry of the compactified 
six-dimensional manifold\footnote{This is a common feature of many models 
dealing with compactified dimensions}. We will see below that this relation 
between $\alpha_3$ and some of the scales of the theory could play an 
important role when building a specific scenario.

We are aware that in an Anisotropic Compactification model
we are re-introducing a possibly large hierarchy in terms of the different
compactification scales. Moreover, the presence of different scales in the model 
suggests that a mechanism stabilizing this hierarchy should be at work \cite{add, infla}.
Anyway, this problem is common to all Anisotropic Compactification models
and its solution is well beyond the motivation of this paper, 
whose principal interest is in a phenomenological description of 
Anisotropic Compactification scenarios. 

In the following, we apply our assumption to three different scenarios
characterized by particularly interesting values of the string scale:
1) $M_s \sim 10^3 \ \gev$, 2) $M_s \sim 10^{11} \ \gev$ and 3) $M_s 
\sim M_{GUT} = 10^{16} \ \gev$.

\subsection{Low-energy String Scenario: TeV-scale Strings}
\label{less}

\begin{figure}[t]
\begin{center}
\epsfig{file=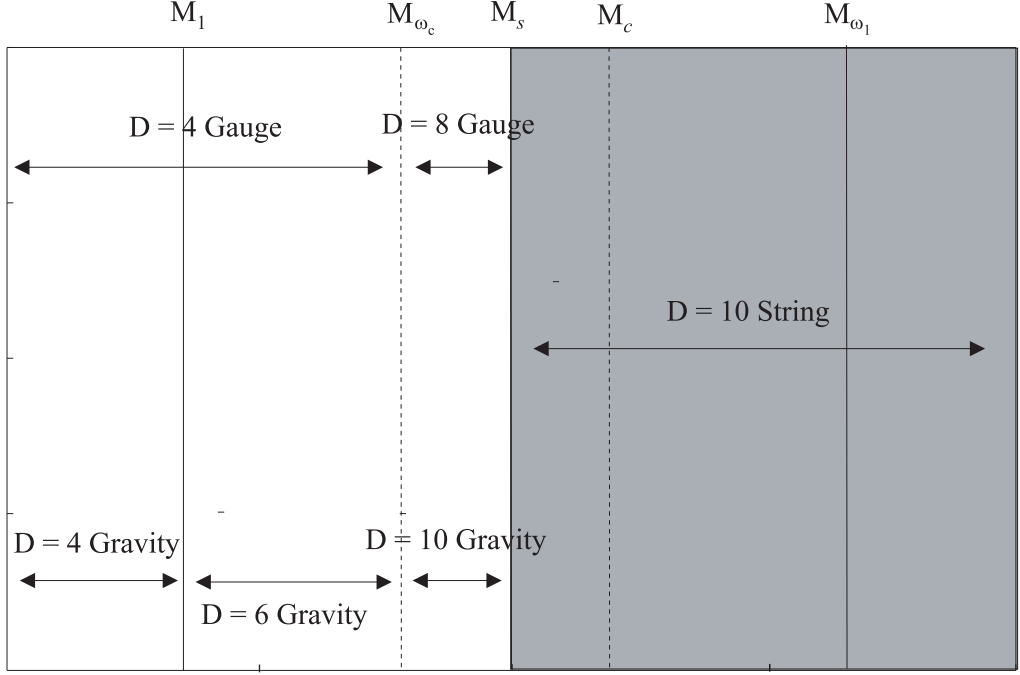,height=9cm} 
\end{center}
\caption{\it \small General pattern of mass scales in scenario (1), where
the lowest scale is the compactification mass $M_1$ and $M_{\omega_1} = M_{Planck}$.}
\label{lowkk}
\end{figure}

In this first regime, we consider a very low string scale, $M_s \sim 10^3 \ 
\gev$. This value is the typical lower bound for new physics beyond the SM 
and has been considered in \cite{add,mmgrav}. Under our Ansatz, this value for 
the string scale results in a fixed value for the largest compactification 
radius, $R_1 = \left ( M_s^2 / M_{Planck} \right )^{-1} \sim 1 \ \mm$, i.e 
$M_1 \sim 10^{-13} \ \gev$. Present gravity experiments have tested Newtonian 
gravity law up to 1 cm, thus not  excluding this value. The scenario is 
qualitatively depicted in Fig.~\ref{lowkk}. If we choose, as a reference 
value, $\alpha_3 / \sqrt{2} = 10^{-2}$, then the complete set of mass scales 
reads:
\bea
M_1 & \sim & 10^{-13} \ \gev \quad < \quad M_{\omega_c} \sim 10^{2} \ \gev 
                      \quad < \quad M_s \sim 10^3 \ \gev \quad < \nn \\
M_c & \sim & 10^4 \ \gev \quad < \quad M_{\omega_1} = M_{Planck} 
                      \sim 10^{19} \ \gev \ . 
\eea
Notice that these values of $R_1$ and $M_s$ drive winding modes with mass $M_{\omega_c}$ 
to be lighter than the string scale, depending on the precise value of $\alpha_3$. 
This, in principle, could have a great phenomenological 
impact and we will discuss it in full detail in sect.~\ref{winLED}.

We recall, however, that such a low value for the string scale could give problems
related to the gauge coupling unification \cite{ckm}. In our scenario, an infinite 
tower of winding modes of the SM particles arise above $M_{\omega_c}$, changing
the logarithmic running of the couplings to a power-law running \cite{ddg}. 
However, in \cite{ross} it was shown that the modified running is not compatible
with the available results for $\alpha_s(M_Z)$ imposing unification at the TeV scale.

As a final comment to the low-energy string scale scenario, we consider the problem of 
SUSY-breaking. If the spontaneous supersymmetry breaking occurs on a distant D$p$-brane 
other than the one where the SM particles are living in, its effect could be
communicated to the visible sector by gravity\footnote{In principle, SUSY-breaking
could be transmitted also by open strings connecting the SM brane and the one 
were SUSY-breaking occurs. This scenario is similar to gauge-mediated models with the 
mass of the messenger proportional to the distance between the two branes.}.
The gravitino mass is, then,
given by:
\be
m_{3/2} \simeq \frac{F}{M_{Planck} } \simeq \frac{M_s^2}{M_{Planck}} \sim 10^{-13} \ \gev \ .
\ee 
where $F$ is the scale of the SUSY-breaking. 
The phenomenological implications of such an extremely light gravitino
have been considered in \cite{fabio}, where also the lower bounds on $m_{3/2}$
are discussed.

The case in which the smallest scale of the model is the winding
mass $M_{\omega_1} = 10^{-13} \ \gev$ is clearly ruled out by experiments. 
Since this extremely light winding mode would interact with SM particles 
experimental signatures of this scenario should have been already discovered. 

\subsection{Middle-energy String Scenario: Planck-Weak scale duality}
\label{mess}
\begin{figure}[t]
\begin{center}
\epsfig{file=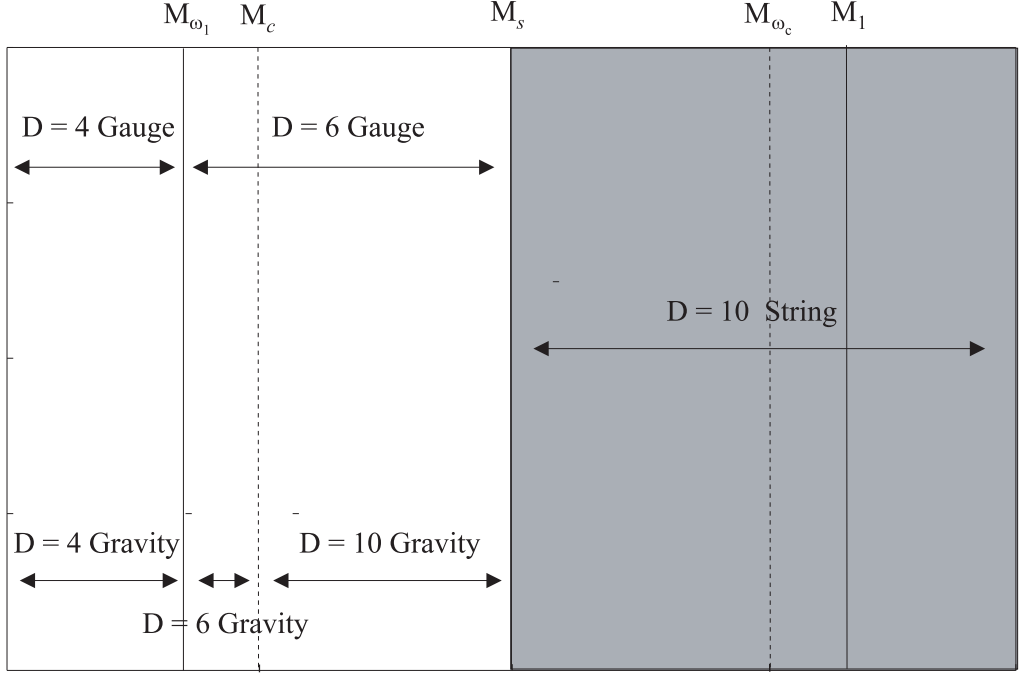,height=9cm} 
\end{center}
\caption{\it \small General pattern of mass scales in scenario (1), where
the lowest scale is the winding mass $M_{\omega_1}$ and $M_1 = M_{Planck}$.}
\label{lowwind}
\end{figure}
In this case we fix the string scale, $M_s$, in a intermediate region 
between the  weak scale and the Planck scale. We choose, as a reference value, 
$M_s = 10^{11}$ Gev. Following our Ansatz, we obtain the two following 
scenarios whenever $M_1$ or $M_{\omega_1}$ are taken as large as 
$M_{Planck}$ and the GUT coupling $\alpha_3 / \sqrt{2} = 10^{-2}$:
\bea
M_1 \sim 10^3 \ \gev \quad & < & \quad M_{\omega_c} \sim 10^{10} \ \gev \quad 
                               < \quad M_s \sim 10^{11} \ \gev \quad \nn \\
& < & 
M_c \sim 10^{12} \ \gev \quad < \quad M_{\omega_1} \sim M_{Planck}=10^{19}
\label{mess1}
\eea
and
\bea 
M_{\omega_1} \sim 10^3 \ \gev \quad & < & \quad M_c \sim 10^{4} \ \gev \quad 
                            < \quad M_s \sim 10^{11} \ \gev \quad  \nn \\
& < & 
M_{\omega_c} \sim 10^{18} \ \gev \quad < \quad M_1 \sim M_{Planck}=10^{19} \ . 
\label{mess2}
\eea
This two scenarios are graphically shown in Fig.~\ref{lowkk} and 
Fig.~\ref{lowwind}, respectively. The effect of growing $\alpha_3$ 
is to reduce the splitting within $M_s - M_c$ and $M_s - M_{\omega_c}$ 
in the first case (see Fig.~\ref{lowkk}) or 
the splitting $M_c - M_{\omega_1}$ and $M_{\omega_c} - M_1$ in the second 
scenario (see Fig.~\ref{lowwind}).

The corresponding Isotropic scenario was considered in \cite{biq}
as a motivation for the large hierarchy between the Planck scale, 
$M_{Planck}$ and the Weak scale, $M_{Weak}$.
In that paper the large hierarchy is just the amplification of the small 
hierarchy
\be
\frac{M_c}{M_s} \sim \alpha_3 \sim \alpha_{GUT} = 10^{-2}
\ee
driven by the $p-6$ exponent of eq.~(\ref{gaugcmi}). 
For this scenario to work, the additional constraint is considered:
\be
M_{Weak} \sim m_{3/2} \sim \frac{M_s^2}{M_{Planck} } ,
\ee 
where the first relation is motivated by the phenomenological 
request of having soft masses at the TeV scale\footnote{
In general, in string theory-derived supergravity models, all
the soft breaking terms are related to the gravitino mass.}.
Under our Ansatz, this ``Weak-Planck duality'' arises 
naturally choosing the smallest scale at TeV.

Soft terms originating from a TeV-compactification scale have been considered
in \cite{quiros}. In principle, the same mechanism can be applied also for 
TeV-winding modes. This can be easily understood via a $T_1$-duality, thus 
exchanging $M_{\omega_1}$ with $M_1$ and replacing the 3-brane with the $5_1$-brane.

\subsection{High-energy String Scenario}
\label{hess}

In this case we consider the string scale $M_s = M_{GUT} = 10^{16} \ \gev$. 
This scenario could be interesting to relate the string scale and 
the GUT unification scale. Following our Ansatz, we obtain the two following 
scenarios whenever $M_1$ or $M_{\omega_1}$ are taken as large as 
$M_{Planck}$ and the GUT coupling $\alpha_3 / \sqrt{2} = 10^{-2}$:
\bea
M_1 \sim 10^{13} \ \gev \quad & < & \quad M_{\omega_c} \sim 10^{15} \ \gev \quad 
                               < \quad M_s \sim 10^{16} \ \gev \quad \nn \\
& < & 
M_c \sim 10^{17} \ \gev \quad < \quad M_{\omega_1} \sim M_{Planck}=10^{19}
\label{hiss1}
\eea
and
\bea
M_{\omega_1} \sim 10^{13} \ \gev \quad & < & \quad M_c \sim 10^{14} \ \gev \quad 
                            < \quad M_s \sim 10^{16} \ \gev \quad  \nn \\
& < & 
M_{\omega_c} \sim 10^{18} \ \gev \quad < \quad M_1 \sim M_{Planck}=10^{19} \ . 
\label{hiss2}
\eea 
From the phenomenological point of view, the Anisotropic Model does not add
any peculiar feature to what already considered in the Isotropic 
scenario of \cite{imr}. Pushing higher the string scale reproduces 
the well known old heterotic scenario of eq.~(\ref{gaughet}), with 
a very small splitting between the different compactification/winding scales.

%% file: tychap3.tex
\section{Winding Modes and Large Extra Dimensions}
\label{winLED}

In the previous section, we stressed that in the framework of a
Type I string theory, many different mass scales arise when 
Anisotropic Compactification of the 6 extra dimensions is considered.
We noticed that quite generally winding modes could appear below the string scale, 
both when the lowest scale is $M_1$ or $M_{\omega_1}$. 
Since winding modes are felt by the gauge interactions is of particular interest to
study if they could give observable signatures at the planned accelerators
and put bounds on their mass from the existing data. These bounds can be derived 
in the context of Type I strings without assuming our Ansatz.

We remind that winding modes can be treated just as KK excitations.
Although the SM particles are confined in 4D, and thus they do not see the opening 
of extra dimensions (and therefore they do not possess any KK mode), 
they still have an infinite tower of massive excitations, 
corresponding to their winding modes. This means that, for example, the present limits
on massive replicas of SM gauge bosons, such as $Z^\prime_{SM}$ or $W^\prime_{SM}$,
can be directly translated into limits on the winding mode mass.
As we stressed in sect.~\ref{type1}, this situation is not at all peculiar to
a model where the SM particles are confined on a 3-brane. If, for example, 
we consider a model with closed strings living in 10D and open strings
tied to a 9-brane, our results for the winding modes can be directly converted 
in results for KK modes, as shown in Tab.~\ref{tabmode}.
We found it easier to restrict ourselves to the case where the SM
is confined on a 3-brane, but we believe useful to stress again that our results
are not at all peculiar to this particular choice. 

There are two possible interesting situations where winding modes could play
a role in the planned high-energy experiments:
\begin{itemize}
\item
The lowest mass scale is $M_{\omega_1} \simeq 1$ TeV.
In this case case, direct searches at present and future accelerators can put
a lower limit on $M_{\omega_1}$, that within our Ansatz translates into a bound on $M_s$.
The phenomenological implications of this scenario are quite similar to 
what was found in \cite{amq}, where the phenomenology of KK modes at the $\tev$-scale 
was studied in the framework of orbifold compactification of heterotic string theory. 
\item
We have a very large compactification radius, $R_1$, and a string scale
around 1 TeV. This case has been qualitatively studied in sect.~\ref{less},
where we pointed out that winding modes slightly lighter than the string scale 
are possible with no need of any particularly strong assumption.
However, since mass scales patterns of this type have been extensively studied
in the recent literature, we believe this scenario deserves a particular attention.
\end{itemize}
In the following, we will briefly remind the basic formulae present in the 
literature and apply them with special attention to the role of the winding modes
in the Large Extra Dimensions scenario.

\subsection{Phenomenology and experimental constraints}
\label{subsecwind}
New gravitational experiments \cite{newexp} could in principle observe 
deviations from the Einstein-Newton 4D gravity due to 
extra dimensions down to a compactification radius $R \sim 10 \ \mim $.
Present gravity experiments test the Newton law down to $R_{exp} \ge 1 \ \cm$.
In \cite{add} a scenario was proposed where new extra dimensions arise at the mm-scale
affecting gravitational interactions, whereas the SM particles are confined 
to a 4D space-time. It was noticed in \cite{aadd} that in 
the framework of Type I strings this is naturally achieved by considering 
the SM living on a 3-brane. 

It is also possible that new high-energy experiments 
at the $\tev$-scale such as NLC or LHC could observe effects due to the emission 
of a graviton into the extra dimensions \cite{grw}.
Although the graviton emission in 4D is suppressed by the 
4D Newton constant, $G_N = 1 / M_{Planck}^2$, the integration 
over all the KK modes of the graviton\footnote{In this picture, gravity 
lives in $(4+n)$-dimensions, and the graviton from the 4-dimensional point 
of view appear as a massless particle with its associated Kaluza-Klein 
modes.} trades this suppression factor with a much smaller one, 
\be
\frac{1}{M_{Planck}^2} \to \frac{E^n}{M_{(D)}^{2 + n} } ,
\ee
where $n$ is the number of extra dimensions accessible to the graviton,
$M_{(D)}$ is the Planck mass in D $= 4 + n $ dimensions and $E$ is the 
c.m. energy. If the D-dimensional Planck mass is at the $\tev$-scale, 
graviton production becomes accessible at the planned accelerators. 
The typical signature will be the production of a SM particle 
and missing energy \cite{grw}. 

The interest of this scenario relies on the fact that the two 
relevant scales $R \sim 1 \ \mm$ and $M_{(D)} \sim 1 \ \tev$ are precisely 
the scales accessible to gravity and high-energy planned experiments, respectively.
Slightly changing these scales dramatically change 
its experimental appeal for the near future. 
In particular, the number of extra dimensions felt by gravity
at the $\tev$-scale and the precise values of $M_{(D)}$ and of the 
compactification radii, $R_i$, are fundamental when making quantitative
predictions on the decay rate into gravitons or on a specific cross-section. 
In the literature, it has been shown that the case with $n=2$ 
(thus, the scenario considered in this paper) gives the favoured signatures 
at the planned experiments.  

By relating the 4D Newton constant $G_N$ with the $D= 4 + n$ one, we get:
\be
\label{eq1}
M^2_{Planck} = 8 \pi R^n M^{n+2}_{(D)}
\ee
where $R$ is the radius of $n$ extra dimensions\footnote{
Recall that $M^2_{(4)} = M^2_{Planck} / 8 \pi = \bar{M}^2_{Planck}$.}.
The Newton constant in Type I strings is also related to the SM gauge coupling,
the string scale $M_s$ and the compactification scales, $M_i$, by
eq.~(\ref{planck}). Combining these two relations, we get 
\bea
M^2_{(6)} &=& \frac{1}{\sqrt{4 \pi} \alpha_3 } M^2_{\omega_c} , 
\qquad M_1 < M_2 = M_3 = M_c \nn \\
M^3_{(8)} &=& \frac{1}{\sqrt{4 \pi} \alpha_3 } M^2_s M_{\omega_1} , 
\qquad M_2 = M_3 = M_c < M_1 \nn \\
M^4_{(10)} &=& \frac{1}{\sqrt{4 \pi} \alpha_3 } M^4_s , 
\label{neweq}
\eea
respectively for $n=2, 4, 6$ Large Extra Dimensions\footnote{
As a simplifying hypothesis, we consider $n$ equally large radii and $6-n$ 
equally small radii.}.
\begin{figure}[t]
\begin{center}
\epsfig{file=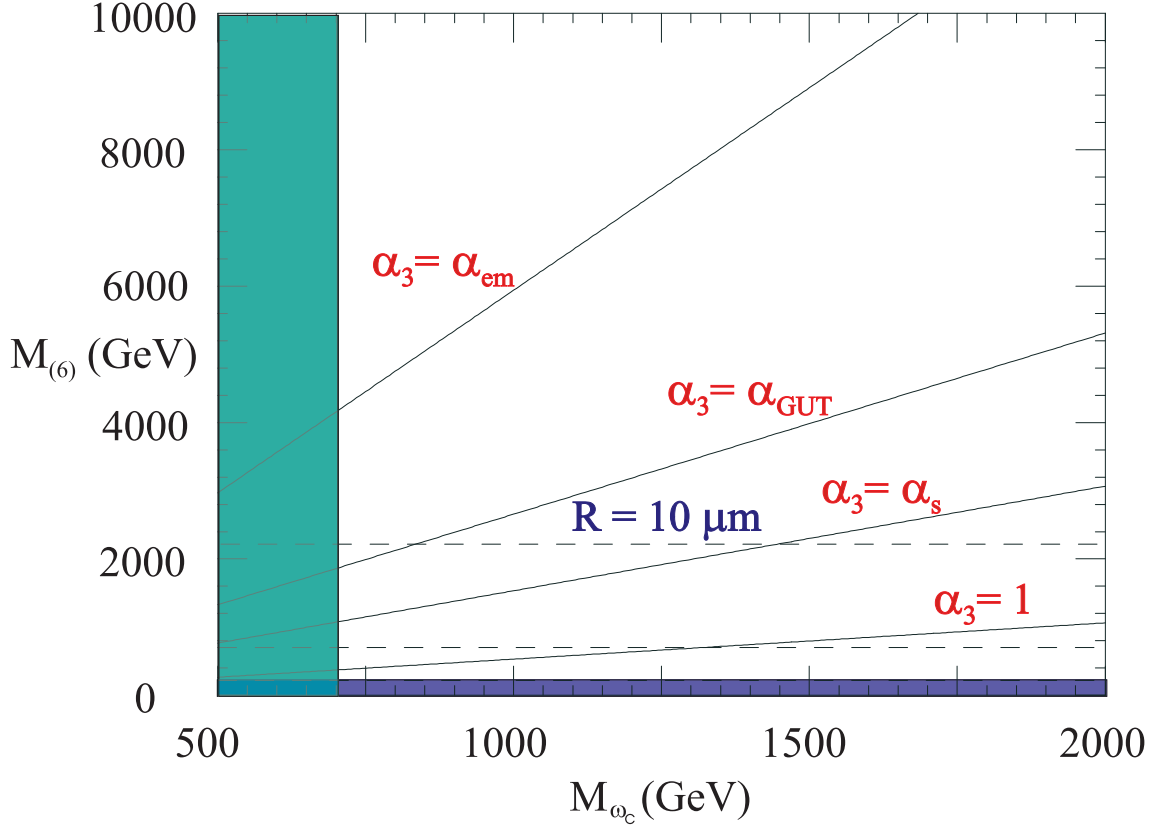,height=11cm} 
\end{center}
\caption{\it \small Bounds on $M_{(6)}$ as a function of 
$M_{\omega_c}$ for different values of $\alpha_3 = \alpha_{em}, \alpha_{GUT}, 
\alpha_s, 1$. The dashed lines are the excluded region if no deviation from 
Newton law is found at $R_1 = 1 \ \mm, 10 \ \mim$. 
The lower shaded area is the excluded region from present gravitational 
experiments testing the Newton law down to $R \sim 1 \ \cm$. The vertical 
left shaded area represent the excluded region from non observation of 
massive replicas of SM bosons.
}
\label{boundm6}
\end{figure}
This means that in the case $n=2$, for which the most promising
experimental signatures are foreseen, the relevant scale in the game
is the winding mode mass $M_{\omega_c}$. This scale is directly related
to the 6-dimensional Planck mass, that is the quantity for which
bounds can be extracted by the experiments (see \cite{grw}). Notice that 
in this particular case the string scale is completely irrelevant from 
the phenomenological point of view and could take any value (above 1 TeV).
However, within our Ansatz the string scale could still be related to 
the winding mode mass: 
\be
M_s^2 = \frac{\sqrt{2}}{\alpha_3} M^2_{\omega_c} = \sqrt{8 \pi} M_{(6)}^2
\ee
and therefore a lower bound on $M_{\omega_c}$ translates into a lower bound for 
$M_s$ also.

By looking to deviations from the Newton law in gravity experiments we
can put direct bounds on $R$, that using eq.~(\ref{eq1}) give limits
on $M_{(6)}$. If we use then Type I strings relations, we get from limits
on $M_{(6)}$ bounds on the winding mode mass and, within our Ansatz, on $M_s$.
In Tab.~\ref{tab2} we resume these bounds for typical values of $R$ 
(with $\alpha_3 = \alpha_{GUT}$).
%
%
\begin{table}[ht]
\centering
\begin{tabular}{||c|c|c|c||}
\hline\hline
     $R_1$  & $M_{(6)}$ & $M_{\omega_c}$ & $M_s$ \\
\hline\hline
1 $\cm$   & $\ge 220 \ \gev$ & $\ge 80  \ \gev$ & $\ge 490 \ \gev$ \\
\hline
1 $\mm$   & $\ge 700 \ \gev$ & $\ge 270 \ \gev$ & $\ge 1.5 \ \tev$ \\
\hline
10 $\mim$ & $\ge 2.2 \ \tev$ & $\ge 840 \ \gev$ & $\ge 5 \ \tev$ \\
\hline \hline
\end{tabular}
\caption{\it{Bounds on $M_{(6)}, M_{\omega_c}$ and $M_s$ from present
and future gravity experiments, with $\alpha_3 = \alpha_{GUT} = 1/24$.
The limit on $M_{(6)}$ is directly derived assuming $4+n$ Newton law.
The limit on $M_{\omega_c}$ comes from the Type I string relations, 
whereas the limit on $M_s$ is obtained using our Ansatz.}}
\label{tab2}
\end{table}
%
%
Present experimental limits on $R_1 \le 1 \ \cm$
put a very weak lower bound to the winding mode mass. 

Existing experimental limits on massive replicas of SM gauge bosons, 
such as $Z'_{SM}$ or $W'_{SM}$, can be directly translated into 
$M_{\omega_c} \ge 700$ GeV \cite{D0CDF}. 
This bound can be used to put limits on $R_1, M_{(6)}$ and $M_s$.
Using the first formula in eq.~(\ref{neweq}), with $\alpha_3 = \alpha_{GUT} = 
1/24$, we obtain the limit $M_{(6)} \ge 1.8$ TeV.
Contextually, using eq.~(\ref{eq1}), one should also have $R \le 0.15 \ \mm$.

\begin{figure}[t]
\begin{center}
\epsfig{file=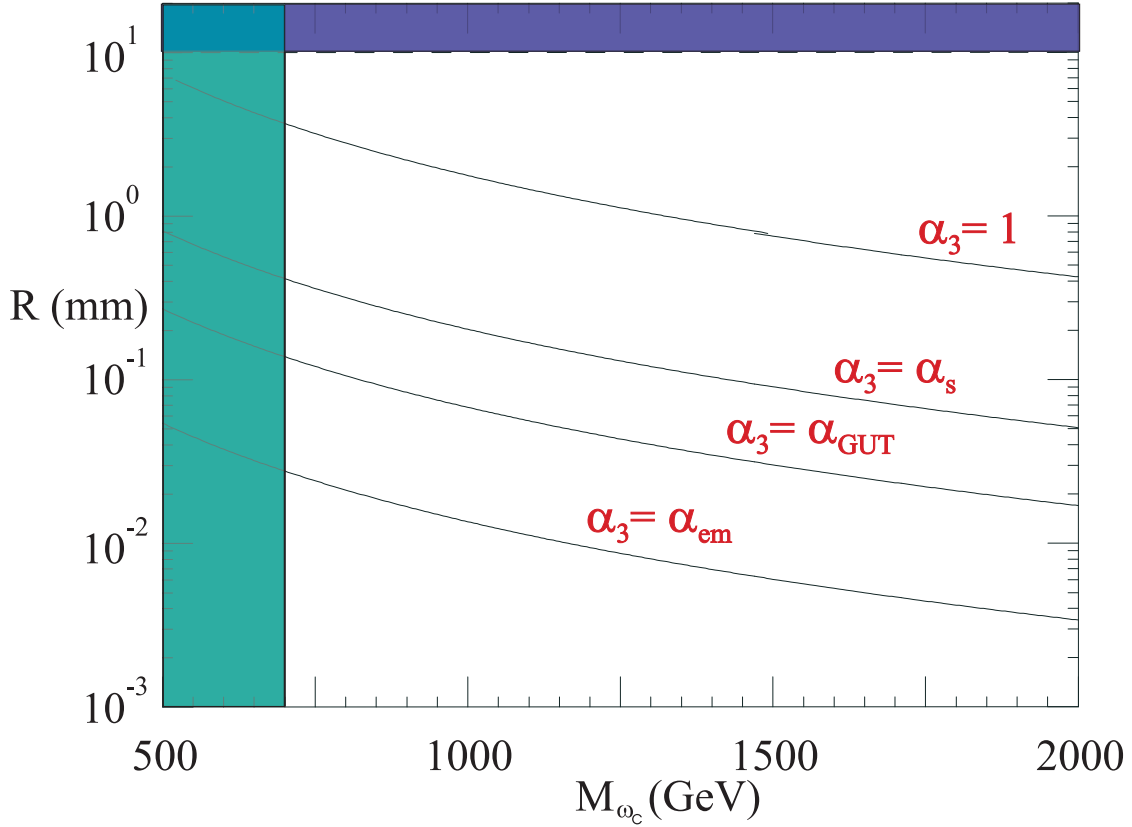,height=11cm} 
\end{center}
\caption{\it \small Bounds on $R_1$ as a function of $M_{\omega_c}$ for 
different values of $\alpha_3 = \alpha_{em}, \alpha_{GUT}, \alpha_s, 1$.
The upper shaded area is the excluded region from present gravitational 
experiments testing the Newton law down to $R \sim 1 \ \cm$. The vertical 
left shaded area represent the excluded region from non observation of 
massive replicas of SM bosons.
}
\label{boundr}
\end{figure}

Non-observation of massive replicas of SM particles at the planned accelerators
would imply even stronger bounds on $M_{(6)}$. For example, non-observation 
at NLC500 (i.e. $M_{Z'_{SM}} \ge 5$ Tev) \cite{NLC}, translates into 
$M_{(6)} \ge 13$ TeV and $R \le 3 \times 10^{-3} \ \mm$.\footnote{Of course 
it is obvious that this bounds could be applied only for $M_{\omega_c} \le M_s$, 
since in the opposite case other new physics effects, such as Regge excitations, 
should be observed.} The dependence of the previous limit on $\alpha_3$ 
is shown in Fig.~\ref{boundm6} and Fig.~\ref{boundr}. 

The Isotropic case $n=6$ provides a direct relationship
between $M_s$ and the experimentally tested $M_{(10)}$.
We face two possibilities:
\begin{itemize}
\item
We observe deviations from the Newton law at the planned gravitational
experiments, $R \sim 1 \ \mm$. In this case, we immediately get 
$M_s \sim 10^{-5} \ \gev$ (for $\alpha_3 = 1/24$). 
Clearly, this situation is excluded by experiments, since the string scale is too low.
\item
$M_s \sim 1 \ \tev$. In this case we get $M_c \sim 10^{-2} \ \gev$
(i.e. $R \sim 10^{-11}$ mm), and thus no extra dimensions could be
observed at the planned gravity experiments. 
Moreover, graviton emission phenomena are suppressed with respect to the $n=2$ case,
due to the high power of $E^6/M^8_{(10)}$ to be compared with $E^2/M^4_{(6)}$.
Although this case is not excluded, its lower phenomenological interest 
with respect to the $n=2$ case is manifest. 
\end{itemize}

Finally, in the $n=4$ case, a limit on $M_{(8)}$ gives a bound
on the combination $M^2_s M_{\omega_c}$. This case is, in a certain 
way, a intermediate situation between the two previous ones. 

If the planned experiments do not discover any winding modes, we
see that we could put bounds on $M_{(6)}$ and on $R$. However, 
we could also imagine a situation where new experiments do discover
some winding modes. If then we have a new accelerator experiment
with c.m. energy above the winding modes threshold, gravity (that feels
both KK and winding modes) start to see new extra dimensions
of radius $R_{\omega_c} = 1 / M_{\omega_c}$. Hence, the Newton constant
is again modified:
\be
M_{(6)}^2 = R^4_{\omega_c} M_{(10)}^6
\ee
If we now make use of eq.~(\ref{neweq}), we get:
\be
M_{(10)}^3 = \sqrt{4 \pi} \alpha_3 M_{(6)}^3
\ee
and thus the suppression factor for graviton emission into the extra dimensions is
\bea
\frac{1}{M_{(4)}^2} & \to & \frac{E^2}{M^4_{(6)}} \nn \\
                    & \to & \frac{E^6}{M^2_{(6)} M^6_{(10)}} \to 
                            \frac{E^6}{4 \pi \alpha_3^2 M^8_{(6)} } \nn
\eea
This means that in this case the suppression factor $E^6/M^8_{(D)}$ typical 
of Isotropic compactification of $n=6$ dimensions gets an
enhancement factor $1/ (4 \pi \alpha_3^2) \simeq 50$ (for $\alpha_3 = 1/24$). 
This enhancement could be relevant when looking for graviton emission 
into extra dimensions above the winding mode threshold. 

If we take into account the present astrophysical and
cosmological bounds on $M_{(6)} \ge 30 \ \tev$, we get:
$R_1 \le 5 \times 10^{-4} \ \mm$, $M_{\omega_c} \ge 12 \ \tev$ and, within our 
Ansatz, $M_s \ge 70 \ \tev$.

%% file: type1.bbl
\begin{thebibliography}{99}
%
\bibitem{dienes}
K. Dienes, Phys. Rept. {\bf 287} (1997) 447.
%
\bibitem{witten}
E. Witten, Nucl. Phys. {\bf B 471} (1996) 135.
%
\bibitem{polwit}
J. Polchinski and E. Witten, Nucl. Phys. {\bf B460} (1996) 525.
%
\bibitem{horwit}
P. Horawa and E. Witten, Nucl.Phys. {\bf B460} (1996) 506.
%
\bibitem{biq}
C. Burgess, L. Ib\'a\~nez and F. Quevedo, hep-ph/9810535.
%
\bibitem{lykken}
J. D. Lykken, Phys. Rev. {\bf D54} (1996) 3693.
\bibitem{add}
N. Arkani-Hamed, S. Dimopoulos and G. Dvali, 
Phys. Lett. {\bf B 429} (1998) 263;\\ 
N. Arkani-Hamed, S. Dimopoulos and G. Dvali, hep-ph/9807344.
%
\bibitem{aadd}
I. Antoniadis, N. Arkani-Hamed, S. Dimopoulos and G. Dvali, 
Phys. Lett. {\bf B 436} (1998) 257; \\
G. Shiu and S. -H. H. Tye, Phys. Rev. {\bf D58} (1998) 106007.
%
\bibitem{grw}
G. Giudice, R. Rattazzi and J. Wells, hep-ph/9811291; \\
E.A. Mirabelli, M. Perelstein and M. E. Peskin, hep-ph/9811337;\\
T. Han, J. D. Lykken and R.J. Zhang, hep-ph/9811350.
%
%
\bibitem{polrev}
For a review see J. Polchinski, hep-th/9611050 and 
C.P. Bachas hep-th/9806199.
%
\bibitem{rev}
For reviews and references see :\\
F. Quevedo, Nucl. Phys. (Proc. Suppl. ) {\bf 62} (1998) 134;\\
F. Quevedo, hep-th/9603074;\\
J. Lykken, Nucl. Phys. (Proc. Suppl. ) {\bf 52 A} (1997) 271;\\
Z. Kakushadze and S.-H. H. Tye, hep-th/9512155;\\
G. Aldazabal, hep-th/9507162;\\
L. Ib\'a\~nez, hep-th/9505098.
%
\bibitem{imr}
L. Ib\'a\~nez, C. Mu\~noz and S. Rigolin, hep-th/9812397.
%
\bibitem{infla}
N. Arkani-Hamed, S. Dimopoulos and J. March-Russell, hep-th/9809124;\\
K.R. Dienes, E. Dudas, A. Gherghetta and A. Riotto, hep-ph/9809406;\\
D. Lyth, hep-ph/9810320; \\
N. Kaloper and A. Linde, hep-th/9811141; \\
G. Dvali and S. -H. H. Tye, hep-ph/9812483.
%
\bibitem{mmgrav}
Z. Kakushadze and S. -H. H. Tye, hep-th/9809147; \\
K. Benakli, hep-ph/9809582;\\
M. Maggiore and A. Riotto, hep-th/9811089;\\
S. Nussinov and R. Shrock, hep-ph/9811323;\\
N. Arkani-Hamed and S.  Dimopoulos, hep-ph/9811353;\\
J. Hewett, hep-ph/9811356;\\
Z. Berezhiani and G. Dvali, hep-ph/9811378;\\
K.R. Dienes, E. Dudas and A. Gherghetta, hep-ph/9811428;\\
N. Arkani-Hamed, S. Dimopoulos and J. March-Russell, hep-ph/9811448;\\
P. Mathews, S. Raychaudhuri and K. Sridhar, hep-ph/9811501; hep-ph/9812486;\\
Z. Kakushadze, hep-th/9812163; \\
A. E. Faraggi and M. Pospelov, hep-ph/9901299.
%
\bibitem{ckm}
E. Caceres, V. S. Kaplunovsky and I. M. Mandelberg, Nucl. Phys. {\bf B 493} (1997) 73.
%
\bibitem{ddg}
T.R. Taylor and G. Veneziano, Phys. Lett. {\bf B 212} (1988) 147;\\
K.R. Dienes, E. Dudas and A. Gherghetta, Phys. Lett. {\bf B 436} (1998) 55;
Nucl. Phys. {\bf B 537} (1999) 47; \\
S. Abel and S. King, hep-ph/9809467.
\bibitem{ross}
G. Ghilencea and G.G Ross, Phys. Lett. {\bf B 442} (1998) 165;\\
T. Kobayashi, J. Kubo, M. Mondragon and G. Zoupanos, hep-ph/9812221.
%
\bibitem{fabio}
A. Brignole, F. Feruglio and F. Zwirner, Nucl. Phys. {\bf B 516} (1998) 13; \\
A. Brignole, F. Feruglio, M. L. Mangano and F. Zwirner, 
Nucl. Phys. {\bf B 526} (1998) 136; \\
A. Brignole, F. Feruglio and F. Zwirner, Phys. Lett. {\bf B 438} (1998) 89.
%
\bibitem{quiros}
I. Antoniadis, S. Dimopoulos, A. Pomarol and M. Quiros, hep-ph/9810410; \\
A. Delgado, A. Pomarol and M. Quiros, hep-ph/9812489.
%
\bibitem{amq}
I. Antoniadis, Phys. Lett. {\bf B 246} (1990) 377; \\
I. Antoniadis, C. Mu\~noz and M. Quir\'os, 
Nucl. Phys. {\bf B 397} (1993) 515;\\
I. Antoniadis, K. Benakli and M. Quir\'os, Phys. Lett. {\bf B 331} (1994) 313.
%
\bibitem{newexp}
J. C. Price, in {\it Proc. Int. Symp. on Experimental Gravitational Physics},
ed. P. F. Michelson, Guangzhou, China (World Scientific, Singapore, 1988); \\
J. C. Price {\em et al.}, NSF proposal, 1996; \\
A. Kapitulnik and T. Kenny, NSF proposal, 1997; \\
J. C. Long, H. W. Chan and J. C. Price, Nucl. Phys. {\bf B 539} (1999) 23.
%
\bibitem{D0CDF}
S. Abachi {\it et al.}, D0 Collaboration, 
Phys. Rev. Lett. {\bf 76} (1996) 3271;\\ 
F. Abe {\it et al.}, CDF Collaboration, 
Phys. Rev. Lett. {\bf 79} (1997) 2192.
%
\bibitem{NLC}
See, for example, A. Leike and S. Riemann, Z. Phys. {\bf C75} (1997) 341.
%
\end{thebibliography}
